\documentclass[preprint,preprintnumbers,showpacs,amsmath,amssymb,aip]{revtex4-1}
\usepackage{graphicx}
\usepackage{dcolumn}
\usepackage{bm}

\begin{document}

\title{Isotropic-medium three-dimensional cloaks for acoustic and electromagnetic waves}

\author{Yaroslav Urzhumov$^*$, Nathan Landy and David R. Smith}

\affiliation{Center for Metamaterials and Integrated Plasmonics,\\
Pratt School of Engineering, Duke University,\\
Durham, North Carolina, 27708, USA\\
$^*$yaroslav.urzhumov@duke.edu}

%

\newcommand{\ba}{\begin{eqnarray}}
\newcommand{\ea}{\end{eqnarray}}
\newcommand{\be}{\begin{equation}}
\newcommand{\ee}{\end{equation}}
\newcommand{\para}{\parallel}

\def \d{\partial}
\def \Re{{\rm Re}}
\def \Im{{\rm Im}}
\def \diag{{\rm diag}}
\def \const{{\rm const}}
\def \eff{{\rm eff}}
\def \sign{{\rm sign}}

\begin{abstract}
We propose a generalization of the two-dimensional eikonal-limit cloak derived from a conformal transformation
to three dimensions. The proposed cloak is a spherical shell composed of only isotropic media;
it operates in the transmission mode and requires no mirror or ground plane.
Unlike the well-known omnidirectional spherical cloaks, it may reduce visibility of an arbitrary object only for a very limited range of observation angles. In the short-wavelength limit, this cloaking structure restores not only the trajectories of incident rays,
but also their phase, which is a necessary ingredient to complete invisibility.
Both scalar-wave (acoustic) and transverse vector-wave (electromagnetic) versions are presented.
\end{abstract}


\maketitle

\section{\label{sec:intro} Introduction}

Transformation optics~\cite{pendry_smith06} (TO) and transformation acoustics~\cite{cummer_njp07,norris_prs08,pendry_li_njp08} (TA)
are the two mathematically similar design methodologies based on the form-invariance of the vector and scalar Helmholtz equation, respectively.
They have recently attracted significant attention as the tools that enabled novel ground-breaking applications, including manipulation of the apparent (effective) width of electromagnetic~\cite{schurig_smith06,kundtz_smith_njp10} (EM)
and acoustic~\cite{chen_chan_apl07,cummer_starr_prl08,zhang_fang11} scatterers. The similarity between the equations describing EM and scalar acoustic waves is so close that many recipes derived for EM applications can be applied in the acoustics domain, and vice versa. In this article, we refer to transformation optics and acoustics collectively as transformation wave dynamics (TWD).

Transformation wave dynamics, in its full generality, employs extremely general coordinate transformations,
including such exotic options as non-bijective (space-folding)~\cite{lai_chan09} and even complex-valued coordinate transformations~\cite{chew_jin_ieee97,popa_cummer_pra11}. However, the more general the choices of transformations are, the more complex the material properties required for their implementation tend to be. For example, non-bijective space-folding transformations require negative-index media~\cite{pendrylens_prl00}, and complex-valued transformations typically lead to polarization-dependent gain~\cite{popa_cummer_pra11}.

In light of this, it is reasonable to ask whether practically useful applications, such as EM or acoustic invisibility, can be accomplished with a more restrictive set of transformations, which lead to a narrower range of material properties. For example, one may wonder whether interesting TWD applications may arise from transformations that require only {\it isotropic} materials. The latter transformations are known to exist at least in two dimensions, where they form a subset of measure zero in the set of all possible real-valued transformations of $\mathcal{R}^2$. Indeed, conformal (angle-preserving) transformations are known to introduce no anisotropy into the material property tensors, a feature that has recently been explored for potential applications in radio engineering~\cite{li_pendry_prl08,liu_smith09}, optics~\cite{ergin_wegener10} and acoustics~\cite{greenleaf_uhlmann08,popa_cummer_prl11,popa_cummer_prb11}.

From the ease of fabrication perspective, one must notice that as long as a TWD device requires exotic medium properties {\it other than anisotropy}, its implementation invariably relies on metamaterials~\cite{smith_apl00}, photonic/phononic crystals~\cite{urzhumov_smith_prl10} or, generally speaking, artificial media with naturally unavailable properties. In the case of {\it optical} cloaking, the other exotic medium property the need for which is more fundamental than anisotropy, is superluminal phase velocity~\cite{urzhumov_smith_prl10}. In fact, all optical cloaking devices proposed so far, including the isotropic conformal cloak of Leonhardt et al.~\cite{leonhardt06,leonhardt_njp06}, rely on the latter property, which in terms of refractive index can be stated as $n<1$. True invisibility devices
--- the ones that can reduce the total scattering cross-section of an object situated in free space ---
preserve not only the amplitude but also the phase of all rays traversing through the device, which inevitably requires that at least somewhere in the cloak, there is a region with $n<1$. The presence of such a region enables a curved ray to have an optical path equal to that of a straight ray in free space.

Since the need for metamaterials in EM cloaking applications is mandated by this exotic property, one may realize that eliminating anisotropy from the implementation does not necessarily lead to a significant reduction in the design complexity, or a significant increase in performance. In fact, metamaterials and photonic crystals can be more easily and efficiently fabricated to have a strong anisotropy than to be perfectly isotropic. Metamaterials in particular are usually made of thin planar elements which provide strong response for one propagation direction only; eliminating the need for omnidirectional response allows one to stack them more densely and achieve higher strength of the response (in one direction only) and lower loss tangents. Thus, at least for two-dimensional optical cloaking applications, reducing the range of transformations to conformal --- and, consequently, the range of materials to isotropic --- does not offer any substantial practical advantages.

The real prospects offered by isotropic-medium designs open up in {\it three dimensions}, where anisotropic media are typically birefringent, that is, they support two normal electromagnetic modes with different dispersion relations. The coupling between the co-propagating modes is so difficult to manage that, in fact, it has been impeding the progress in three-dimensional transformation optics in the past five years.
In the domain of elastodynamics (acoustics of elastic media), anisotropy leads to strong coupling~\cite{urzhumov_smith_njp10} between longitudinal pressure ($p$) and transverse shear ($s$) waves, leading to issues similar to birefringence in electrodynamics.

Unfortunately, conformal transformations in three dimensions are virtually non-existent, with one exception of the so-called sphere inversion transformation given by the famous Liouville's theorem~\cite{blair00}. For that reason, the main focus of the recent conformal TO work has been on finding transformations that are approximately conformal~\cite{li_pendry_prl08,ergin_wegener10} and thus lead to materials with anisotropy so weak that it can be neglected in the design, and then partially compensated by numerical optimization~\cite{landy_kundtz_smith10}. This branch of TO is now known as quasi-conformal TO (QCTO)~\cite{li_pendry_prl08,landy_kundtz_smith10}.

Eliminating anisotropy from the TWD designs, however, imposes serious --- and quite fundamental --- limitations to the performance of the resulting TWD devices. In particular, it is widely believed that a true, omnidirectional invisibility device
cannot be obtained using only isotropic media, a belief articulated in the works of Leonhardt~\cite{leonhardt06} and Greenleaf et al.~\cite{greenleaf_phys03}. In this paper, we show that although cloaking devices based on conformal transformations are unlikely to become omnidirectional, for one specific angle of incidence they can indeed restore the amplitude and phase of the incident waves.

This paper is organized as follows: first, we analyze the performance of the two-dimensional conformal cloak from Refs.~\cite{leonhardt06,leonhardt_njp06} using ray-tracing simulations that include the phase information, as well as full-wave simulations, and show that it can be made to perform as a unidirectional invisibility device. Subsequently, we generalize that design to three-dimensions and propose an all-isotropic unidirectional cloak of a spherical shape for acoustic and electromagnetic applications. The EM (vector-wave) version of the proposed cloak is polarization-insensitive.

\section{\label{sec:2d} Analysis of the two-dimensional conformal cloak}

\begin{figure}
\centering
\begin{tabular}{cc}
\includegraphics[width=0.5\columnwidth]{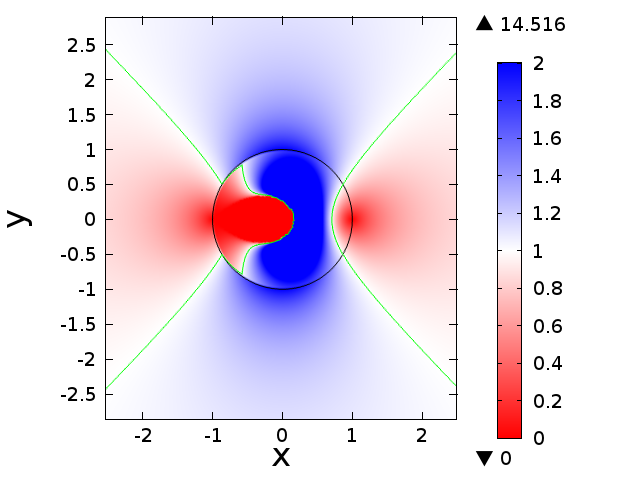}&
\includegraphics[width=0.5\columnwidth]{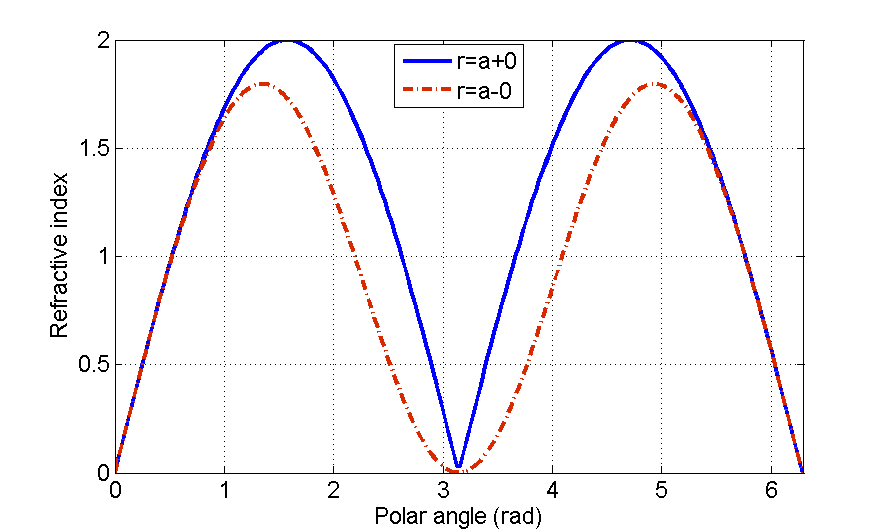}\\
(a)&(b)\\
\includegraphics[width=0.5\columnwidth]{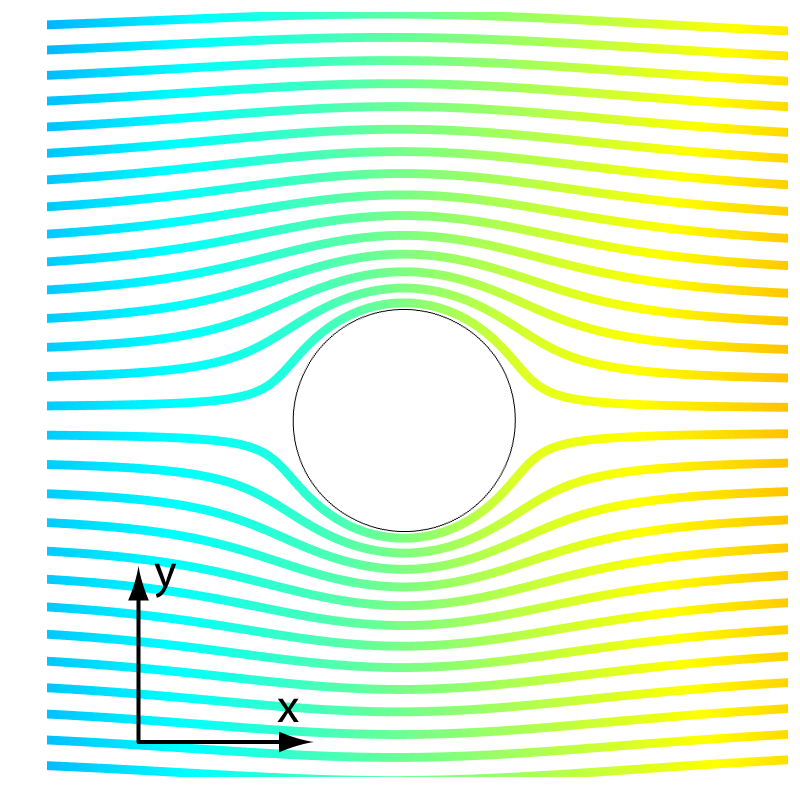}&
\includegraphics[width=0.5\columnwidth]{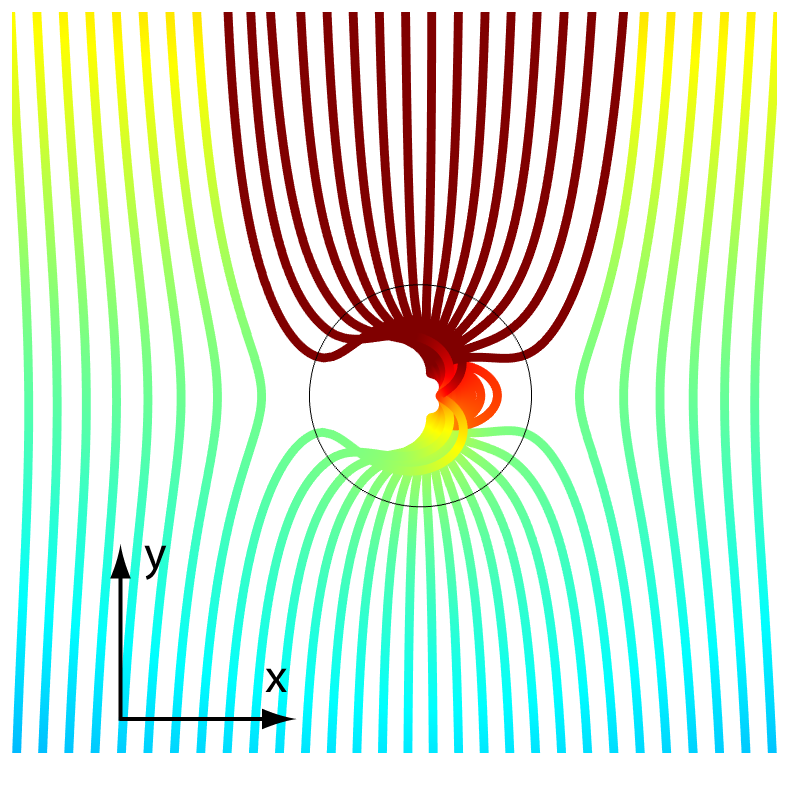}\\
(c)&(d)
\end{tabular}
\caption{(color online). Two-dimensional conformal cloak proposed in Refs.~\cite{leonhardt06,leonhardt_njp06}:
(a) refractive index profile, with the boundary between superluminal ($n<1$) and sub-luminal quadrants shown by green lines;
the color shade inside the black circle ($r=a$) is cut off at refractive index values $n=0$ and $n=2$, respectively;
(b) refractive index as a function of the polar angle on the two sides of the discontinuity at $r=a$;
(c) ray-tracing simulation showing trajectories and phase (color online) of the rays in the eikonal limit,
assuming a plane wave incident along the $x$ direction;
(d) same as (c) with a plane wave propagating in the $y$ direction.
}
\label{fig:conformal_raytracing}
\end{figure}

Cloaking devices that arose from the TWD methodology are derived from coordinate transformations that compress certain regions of physical space into a zero-volume region in the virtual space. The two-dimensional variant of the cloak proposed by Pendry et al.~\cite{pendry_smith06,schurig_smith06} shrinks a cylindrical volume isotropically to an infinitesimal diameter cylinder, a transformation that requires anisotropic material properties but leads to a perfectly cylindrically symmetric distribution thereof. The inherently two-dimensional conformal cloak of Leonhardt et al.~\cite{leonhardt06,leonhardt_njp06} flattens a cylinder of radius $a$ to a flat sheet, which can be accomplished using a conformal map of an entire two-dimensional plane. Such conformal maps, known as Zhukovsky transform in hydrodynamics, have been applied to electrodynamic wave propagation problems long before~\cite{meinke62} the discovery of electromagnetic invisibility. 

Specifically, the transformation used here and in Refs.~\cite{leonhardt06,leonhardt_njp06} maps the exterior region $r>a$ of a cylinder onto the entire plane with excluded line segment $x\in[-a,a]$. The orientation of this segment, which we refer to as the principal axis of the conformal cloak (denoted $x$ here), sets the direction of incidence for which this device may reduce visibility of the cylindrical cavity bounded by the circle $r=a$. In complex variables, this transformation can be written as~\cite{leonhardt06,leonhardt_njp06}
\be
w(z)=z+a^2/z,
\label{eq:conformal_map}
\ee
where $z=x+iy$ is the complex-valued representation of two-dimensional Cartesian coordinates, and $w$ is the complex-valued coordinate in the virtual space. The practical result is that in the physical space, the refractive index distribution has to be
\be
n_e(x,y)=\left| 1-\frac{a^2}{(x+iy)^2} \right| = \sqrt{1+\frac{a^4-2a^2r^2\cos2\phi}{r^4}},
\label{eq:n_conformal_exterior}
\ee
where in the second expression we have used the polar coordinates $(r,\phi)$ oriented such that the positive $x$ axis corresponds to $\phi=0$.
This transformation is to be used only in the region exterior to the cylinder ($r>a$); what happens inside of it will be discussed below.

The key feature of the refractive index distribution (\ref{eq:n_conformal_exterior}) is that $0\le n_e \le 2$ for all $r>a$; the regions with superluminal phase velocity ($n<1$) occupy two quadrants with roughly 50\% of the cloak volume, as shown by the isocontour $n=1$ in Fig.~\ref{fig:conformal_raytracing}(a). As discussed above, the presence of such regions is fundamental to the operation of any true invisibility device, as those regions enable optical path compensation.

\begin{figure}
\centering
\begin{tabular}{cc}
\includegraphics[width=0.5\columnwidth]{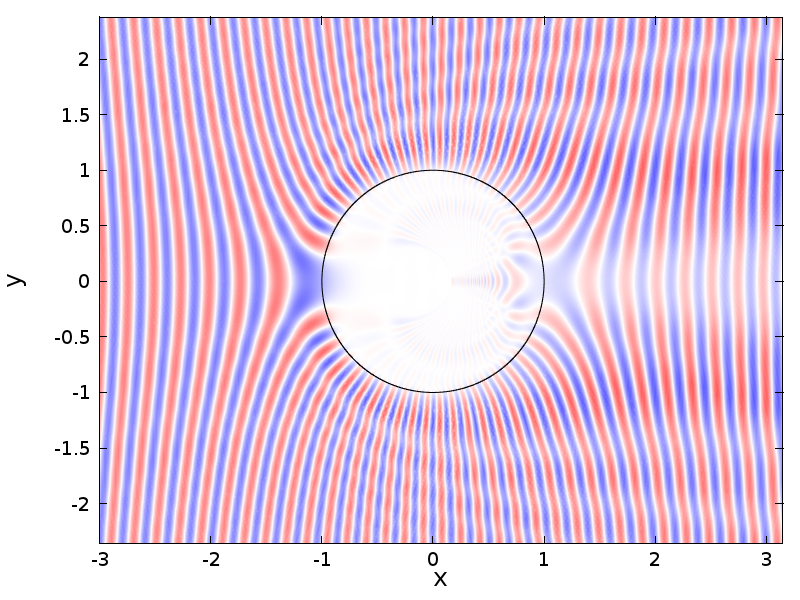}&
\includegraphics[width=0.5\columnwidth]{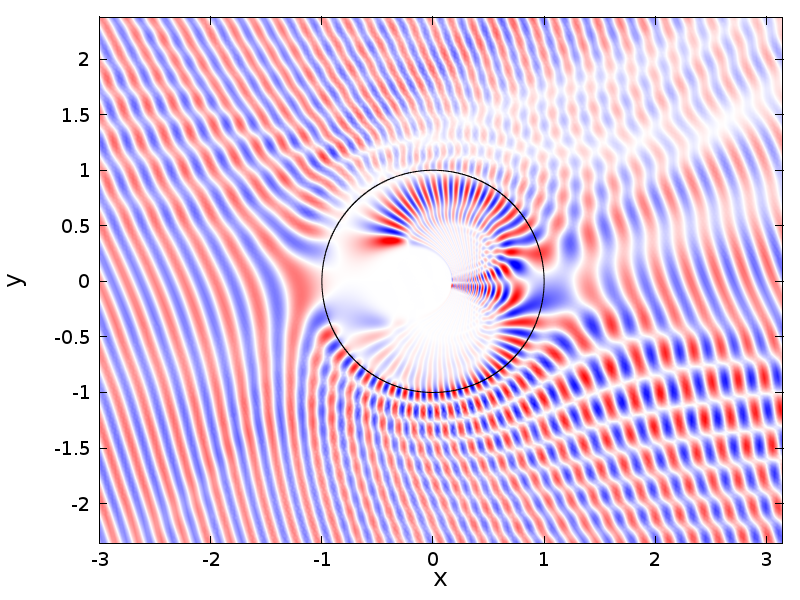}\\
(a)&(b)\\
\includegraphics[width=0.5\columnwidth]{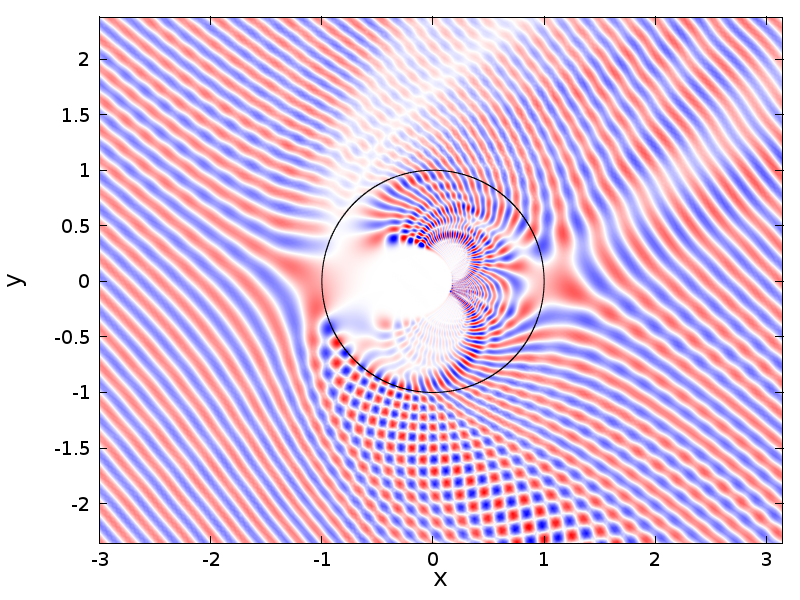}&
\includegraphics[width=0.5\columnwidth]{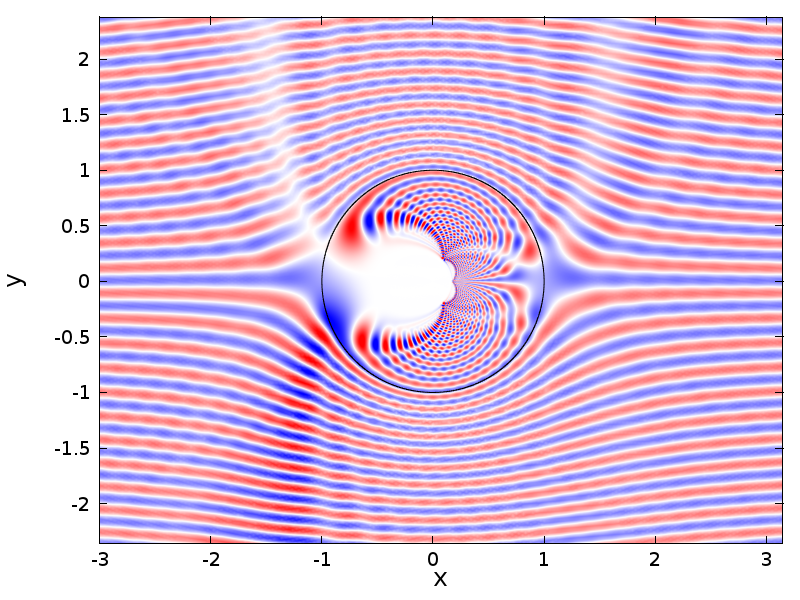}\\
(c)&(d)\\
\end{tabular}
\caption{(color online). Two-dimensional conformal cloak proposed by Leonhardt et al.~\cite{leonhardt06,leonhardt_njp06}.
Full-wave simulations of the structure shown in Fig.~\ref{fig:conformal_raytracing}.
Angle of incidence: (a) 0, (b) $\pi/8$, (c) $\pi/4$, (d) $\pi/2$. TE polarization ($E$ field out of plane) is assumed. Refractive index distribution is implemented using $\epsilon_z=n^2$, with $n$ given by Equations~(\ref{eq:n_conformal_exterior},\ref{eq:n_conformal_interior}) and in-plane permeability $\mu=1$. Free-space wavelength $\lambda_0=a/5$.
}
\label{fig:conformal_fullwave}
\end{figure}

The main idea behind the device (\ref{eq:n_conformal_exterior}) is that a cylinder appears as a flat sheet, which --- at least in the limit of geometrical optics --- is invisible in the direction of the $x$-axis. Its low visibility for this particular angle of incidence is confirmed with the ray-tracing (Fig.~\ref{fig:conformal_raytracing}(c)) as well as full-wave (Fig.~\ref{fig:conformal_fullwave}(a)) simulations performed with a commercial EM solver COMSOL. Of course, for any incidence angle other than $\phi=0$, the flattened sheet is not expected to be invisible, which makes the device given by the profile (\ref{eq:n_conformal_exterior}) a unidirectional invisibility device.

To increase the range of incidence angles for which the conformal mapping device (\ref{eq:n_conformal_exterior}) can be invisible, it was proposed to extend the transformation onto the second Riemann sheet of the transformation (\ref{eq:conformal_map}) and ensure --- by theoretical means~\cite{leonhardt_tyc_science09,chen_tyc11,perczel_leonhardt11} other than transformation optics --- that all rays emerging on the second Riemann sheet follow closed orbits. In order to implement this behavior, the interior of the circle $r=a$ would have to be filled with inhomogeneous spatial refractive index distribution; for example, the Hooke's profile proposed in Refs.~\cite{leonhardt_njp06,chen_tyc11} gives the following refractive index for $r<a$:
\be
n(x,y)=n_e(x,y)\sqrt{1-\frac{|w(z)-2a|^2}{(4a)^2}},
\label{eq:n_conformal_interior}
\ee
where $n_e$ is given by the same function (\ref{eq:n_conformal_exterior}), and $w(z)$ --- by the same equation (\ref{eq:conformal_map}). The complete index profile is shown in Fig.~\ref{fig:conformal_raytracing}(a).

Notably, this index distribution is discontinuous across the boundary $r=a$, as illustrated by Fig.~\ref{fig:conformal_raytracing}(b).
That such a discontinuity is unavoidable in the design methodology proposed in Refs.~\cite{leonhardt06,leonhardt_njp06} can be easily understood using the theory of functions of complex variables. One form of the uniqueness theorem for analytical functions of a complex variable states that an analytical function is uniquely determined by its values on a finite-length interval; the values of the function can be determined unambiguously using the process known as analytic continuation. Thus, if the refractive index $n'(w)$ in the virtual space has to be continuous across the branch cut $\Re w=[-2a, 2a]$, it has to be given by the same analytic function on both Riemann sheets that are connected by that branch cut. Choosing $n'(w)\equiv 1$ on one Riemann sheet forces one to either also pick $n'(w)\equiv 1$ on the other sheet, or to have a discontinuity of refractive index.

The behavior of the structure with refractive index (\ref{eq:n_conformal_interior}) filling the interior of the circle $r=a$ in physical space is modeled using ray-tracing and full-wave simulations. To make a fair assessment of this structure, which is expected to work as a cloak only in the eikonal (short-wavelength) limit, we have chosen the free-space wavelength $\lambda_0=a/5\ll a$ for the full-wave simulations, such that the conditions of geometrical optics are satisfied. Note that in the earlier assessment of this structure, the wavelength was chosen to exceed not only the diameter $d$ of a cloaked object ($d<a$), but even the diameter of the entire cavity ($2a$); in that regime, the cavity and anything in it is poorly visible to begin with and thus its poor visibility is not sufficient evidence of the cloaking performance.

While the ray-tracing simulations suggest that the full index profile (\ref{eq:n_conformal_exterior},\ref{eq:n_conformal_interior}) does indeed allow transmitted rays to return to their original trajectories after passing through the cloak, it is obvious that the transmitted rays leave the cavity ($r<a$) with incorrect field amplitude and phase~\cite{chen_tyc11}. The phase delay is introduced to every ray that spends any amount of time inside the circle $r=a$, whereas the amplitude deviations are caused by the partial reflections on the refractive index discontinuity.
These partial reflections are insignificant for the incidence angle $\phi=0$, at which the circle $r=a$ is perfectly cloaked, but they become the limiting factor for all other incidence angles, at which the rays do hit the surface $r=a$ from which they can get reflected.
Note that the ray-tracing simulations presented here as well as in the original works~\cite{leonhardt_njp06,chen_tyc11} completely disregard reflected rays, making an erroneous impression that the transmission through this structure can be close to perfect. Strong reflectivity is revealed by the full-wave models shown in Fig.~\ref{fig:conformal_raytracing}(c,d).

To summarize, for any angle of incidence other than $\phi=0$, the device composed of index profiles (\ref{eq:n_conformal_exterior}) and (\ref{eq:n_conformal_interior}) has a wide shadow, whose width at the $90^\circ$ incidence is roughly twice the diameter of the circle ($=2a$) and about four times the diameter of the cloaked cavity --- the region of space not probed by any of the incident rays. This behavior is also observed at other angles of incidence, as evident from the full-wave simulations in Fig.~\ref{fig:conformal_fullwave}.

\section{\label{sec:3d_acoustic}
Three-dimensional isotropic-medium unidirectional acoustic cloak}

The analysis presented in Section~\ref{sec:2d} suggests that although isotropic metamaterial cloaks proposed in the earlier works can hardly be used as omnidirectional invisibility devices, for one angle of incidence they do have the potential to operate as cloaks. In two dimensions, they require the same type of hard-to-achieve response as their omnidirectional, anisotropic counterparts, namely refractive index in the $0-1$ range with very small imaginary part, and therefore are unlikely to beat the performance of the latter. However, extending these solutions into three dimensions would offer advantages such as elimination of birefringence, and, for acoustic applications, also reducing the coupling between longitudinal (pressure) and transverse (shear) waves.

\begin{figure}
\centering
\begin{tabular}{cc}
\includegraphics[width=0.45\columnwidth]{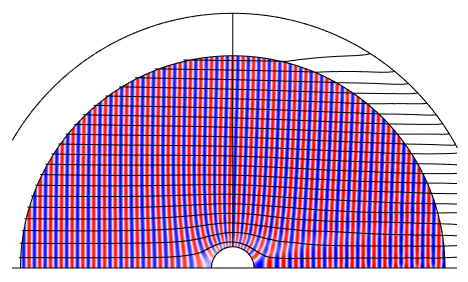}&
\includegraphics[width=0.45\columnwidth]{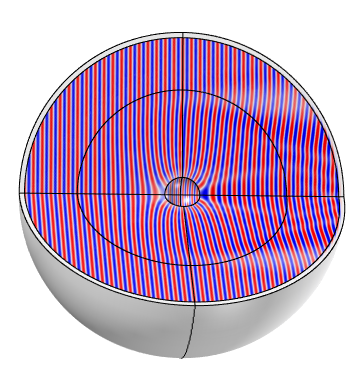}\\
(a)&(b)\\
\end{tabular}
\caption{(color online). Three-dimensional unidirectional acoustic cloak consisting of isotropic medium:
(a) pressure distribution on the cross-section of the cloak. Black lines show the streamlines of acoustic flux;
acoustic wavelength at $r\gg a$ is $\lambda_0=a/2$;
(b) three-dimensional picture of the cloak with a finite exterior radius $R_{cut}=4a$;
wavelength in ambient medium (at $r>R_{cut}$) $\lambda_0=a/2$;
the index transition is smoothed on a spatial scale $\Delta R=a/5$.
Acoustic wave impedance is assumed constant in both simulations; the speed of sound varies according to $c_s=c_0/n$,
where $n$ is prescribed by the revolution of profile (\ref{eq:n_conformal_exterior}).
}
\label{fig:acoustic}
\end{figure}

Three-dimensional extensions of two-dimensional TO and TA designs are difficult to come up with if the device has to be a precise implementation of a TO/TA prescription. However, in the so-called eikonal limit one only manages the trajectories and phases of the rays, allowing the impedance to vary almost arbitrarily in the device. Avoiding sharp discontinuities of the wave impedance allows one to achieve performance in eikonal-limit TO devices, including cloaks, that approaches that of the exact TO design.

In the eikonal limit, a three-dimensional generalization of a TWD design can be obtained by a simple revolution of the refractive index distribution along some axis. For example, it is well-known that the refractive index profile of the omnidirectional 3D cloak of Pendry et al. is the same on any cross-section passing through its center as is the index profile of the 2D (cylindrical) cloak. Likewise, the ground-plane cloak that hides objects in the reflection mode can be quite easily generalized into 3D by rotation around a line normal to the ground plane. Here, we propose revolving the conformal cloak studied in Section~\ref{sec:2d} around its principal axis, i.e. around the direction from which it can be actually invisible.

As a result of the revolution, the cloaked cavity becomes spherical in shape, as seen from Fig.~\ref{fig:acoustic}.
Acoustic frequency-domain simulations based on the scalar Helmholtz equations can take advantage of the cylindrical symmetry of the problem
using a standard axisymmetric acoustic solver~\cite{comsol}.
The boundary condition on the surface of the spherical cavity is assumed to be Neumann ($\partial p/\partial n=0$),
which corresponds to an acoustically hard scatterer, such as a solid object with a high bulk modulus ($K$) situated in a fluid much smaller bulk modulus ($K_0\ll K$). Very similar results (not shown) were obtained with Dirichlet boundary condition ($p=0$), which is often used in acoustics to approximate an interface between two fluids. These two boundary conditions are also known as high- and low-impedance boundaries, respectively. As discussed in Ref.~\cite{urzhumov_pendry11}, for three-dimensional cloaks there is usually little difference between these two boundary conditions, since the surface is compressed and appears either as a point (for omnidirectional cloaks) or as a line (for the directional cloak described here) in the virtual space. In either case, point and line objects do not have a boundary condition associated with them, which explains why the structure is insensitive to the choice of the boundary condition in the physical space.

Note however that allowing the waves to leak into the cavity by using a continuity boundary condition still distorts the wave pattern and leads to additional scattering, which is evident from Fig.~\ref{fig:conformal_fullwave}(a), where the interior of the cavity is filled with a refractive index and the surface $r=a$ is modeled with a continuity boundary condition. As it turns out, in both 2D and 3D the best performance is achieved when the surface of the cavity ($r=a$) is coated with a perfect reflector, as opposed to filling the cavity with the refractive index prescribed by (\ref{eq:n_conformal_interior}).

So far we have been neglecting the fact that the refractive index distribution (\ref{eq:n_conformal_exterior}) is filling the entire space, making this design impractical as an invisibility device. The simulation in Fig.~\ref{fig:acoustic}(a), which shows nearly ideal cloaking, assumes that refractive index extends to infinite radius, and an artificial boundary (spherical PML) is introduced to mimic an infinitely extended cloak.
A prior study~\cite{urzhumov_pendry11} showed that simply cutting off the transformation at a finite radius $R_{cut}$ results in a formation of a strong shadow generated by the refractive index mismatch $\delta n$ at $r=R_{cut}$. Since the index given by (\ref{eq:n_conformal_exterior}) scales roughly as $1+\alpha/r$ at $r \gg a$, the product $\delta n R_{cut}$ does not converge to zero, and the effective scattering width of the cut-off cloak does not converge to zero in the limit $R_{cut}\gg a$. This type of scattering can be reduced, although not completely eliminated, by introducing a continuous transition from the index profile (\ref{eq:n_conformal_exterior}) to the homogeneous index of the ambient medium, which is normalized to unity in our calculations. The resulting simulation of a cloak with finite exterior radius is shown in Fig.~\ref{fig:acoustic}(b).

To summarize, it is possible to achieve directional acoustic invisibility for a situation where the positions of the source and the detector relative to the cloaked object are known, using continuous distributions of isotropic density and isotropic bulk modulus. Such acoustic media are readily available as natural materials, random composites or mixtures, as well as ordered/structured metamaterials~\cite{chen_chan_apl07,cummer_starr_prl08,popa_cummer_prl11,popa_cummer_prb11,zhang_fang11}.
Directional acoustic cloaks still require media with the speed of sound ($c_s=c_0/n$, where $c_0$ is the speed of sound in ambient medium and $n$ is the refractive index in the cloak) as high as possible~\cite{urzhumov_smith_njp10}, and their performance is still limited by the highest achievable value of $c_s$. We emphasize though that for limited-bandwidth cloaking applications only the phase velocity of sound needs to be high enough, and no statement about the group velocity is needed for such applications.
In acoustic metamaterials, at a select frequency the phase velocity of sound can be engineered to be extremely high --- much higher than the speed of sound estimated from the quasistatic bulk modulus and density of the materials involved, which opens up the possibility for nearly perfect acoustic cloaking~\cite{chen_chan_apl07,cummer_starr_prl08,zhang_fang11}.

\section{\label{sec:3d_EM}
Three-dimensional polarization-insensitive electromagnetic cloak}

In the remainder of this paper, we apply the cylindrically-symmetric refractive index profile described in Section~\ref{sec:3d_acoustic} to the problem of electromagnetic cloaking. As mentioned before, eliminating anisotropy from three-dimensional designs does away with birefringence, and allows a simple transition of TO solutions to their eikonal-limit.

\begin{figure}
\centering
\begin{tabular}{cc}
\includegraphics[width=0.45\columnwidth]{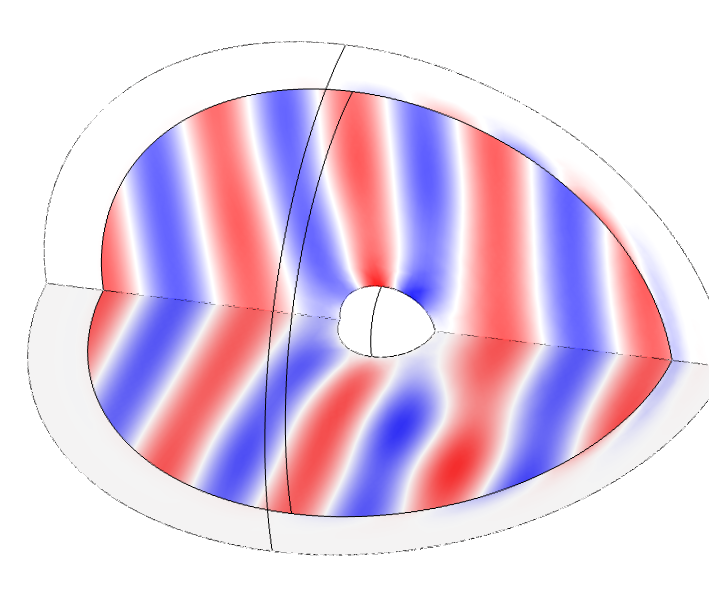}&
\includegraphics[width=0.45\columnwidth]{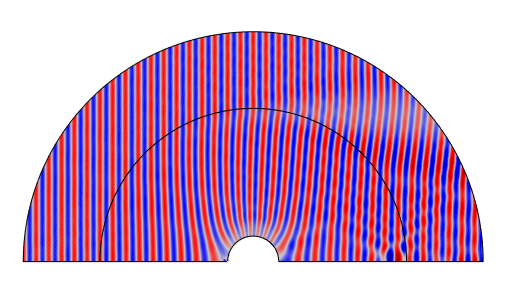}\\
(a)&(b)\\
\end{tabular}
\caption{(color online). Three-dimensional polarization-insensitive electromagnetic cloak made of isotropic non-magnetic medium:
(a) three-dimensional full-wave simulation with free-space wavelength $\lambda_0=0.8a$
showing electric field component transverse to the propagation direction;
simulation domain (excluding the PML) radius $R=4a$;
(b) 2.5-dimensional model with $\lambda_0=a/2$ and cloak exterior radius $R_{cut}=6a$; transition width $\Delta R=0.3a$.
}
\label{fig:EM}
\end{figure}

In order to stay within the validity of eikonal approximation, the cloaked object and the cloak itself must be many wavelengths in diameter.
Modeling optically large structures with inhomogeneous refractive index tends to be a very computationally intensive task, and full-wave simulations of three-dimensional cloaks and other TO structures spanning many wavelengths are still not abundant in the literature. For a cylindrically-symmetric structure, one can reduce the computational domain to one quarter of the full domain by using two orthogonal symmetry planes intersecting along the axis of revolution. However, even with that reduction, Finite Element simulations still often require prohibitively large resources and CPU time; the largest domain diameter to wavelength ratio we could efficiently model in this fashion was about $8$, as shown in Fig.~\ref{fig:EM}(a).

Fig.~\ref{fig:EM}(a) reports a full-wave simulation of the directional cloak made of isotropic graded-index medium with the refractive index distribution obtained by revolution of (\ref{eq:n_conformal_exterior}) around the $x$ axis. Our particular implementation of this refractive index in that simulation assumes a dielectric-only medium; similar results can be obtained with magnetic-only medium or a dielectric-and-magnetic medium whose refractive index satisfies the same equation.

From the TO perspective, the described structure operates by compressing a spherical cavity covered with a perfect reflector (either PEC or PMC) to a zero-diameter, finite-length ``needle". The needle is almost invisible from its tip in the geometrical optics limit, and it scatters little even when the wavelength is finite. Importantly, since the structure is rotationally-invariant with respect to the needle direction, the scattering pattern does not depend on the polarization of transverse EM waves incident upon it. As a result, this structure works as a directional cloak for an arbitrary polarization of incident light, including elliptic and partial polarizations.

In order to characterize its scattering pattern deeply in the geometrical optics (eikonal) regime, we have developed a quasi-two-dimensional modeling method for vector EM fields, which can be referred to as {\it 2.5D modeling}. The method is an extension of the well-known axisymmetric modeling technique used above for the acoustic wave cloak. The standard axisymmetric modeling assumes that all fields are independent of the azimuthal angle; therefore, it cannot be applied to vector EM fields, whose transverse polarization breaks down the cylindrical symmetry even when the wave vector is aligned with the revolution axis of the structure. It is well known that EM scattering on a sphere (Mie scattering) has a distinct dipolar azimuthal dependence of the scattered field. The mathematical origin of this dipolar dependence can be seen by projecting the electric field vector of a plane EM wave onto the basis vectors of cylindrical coordinates: although the Cartesian component $E_x$ is independent of the azimuthal angle $\phi$, the cylindrical components $E_\phi=E_x\cos\phi$ and $E_r=E_x\sin\phi$ both have dipolar angular dependence.

The 2.5D method differs from the usual axisymmetric modeling in that it includes a prescribed variation of the fields as a function of the azimuthal angle. In linear electrodynamics, the fields can be decomposed into the Fourier series,
\be
 E_{r,z,\phi}(r,z,\phi) = \sum_m \tilde E_{r,z,\phi}^{(m)}(r,z) e^{im\phi}
\ee
and each cylindrical harmonic propagates independently in the structure whose material properties are $\phi$-independent.
Note that in general all three field components are non-zero, and the full vector Helmholtz equation including the three unknown field components $E_{r,z,\phi}^{(m)}(r,z)$ must be solved on the two-dimensional cross-section of the simulation domain for each $m$. After eliminating the $\phi$-dependence, the equation depends on the azimuthal number $m$, and has to be solved for each cylindrical harmonic present in the incident field. Fortunately, if we are interested in modeling wave propagation at one angle of incidence (along the axis of revolution), only $m=\pm 1$ harmonics are non-zero in the incident field, so the equation needs to be solved only twice. In fact, the $m=+1$ and $m=-1$ harmonics are related by complex conjugation, which further reduces the computational load by a factor of two.

The results for the directional three-dimensional cloak with exterior radius $R_{cut}=6a$ are shown in Fig.~\ref{fig:EM}(b), where the free-space wavelength is chosen to be $\lambda_0=a/2$. The electric field pattern is very similar to the pressure pattern observed in the acoustic cloak shown in Fig.~\ref{fig:acoustic}(b). In this simulation, we have used incident electric field polarized in the $\phi=0$ direction; due to rotational invariance of the refractive index distribution, it is sufficient to demonstrate cloaking for one polarization only, and the result extrapolates trivially to other linear polarizations, as well as their superpositions (elliptic polarizations).

\section{\label{sec:conclusion} Conclusions}

In conclusion, we have analyzed the difficulties associated with the proposals to use isotropic media and isotropic metamaterials for cloaking applications.  We have explained that achieving omnidirectional cloaking using isotropic-only media is extremely difficult
even in two dimensions, where the design process can be guided by conformal transformation optics.
Although no clear path to any such design has been identified so far,
lifting the requirement of all-angle invisibility allows one to arrive at useful two- and three-dimensional cloaking structures for acoustic and electromagnetic applications. Directional three-dimensional isotropic-medium cloaking designs eliminate the need to manage wave birefringence and should therefore be substantially easier to make than their anisotropic counterparts.

\section*{\label{sec:ack} Acknowledgements}

This work was supported by the U.S. Navy through a subcontract with SensorMetrix (Contract No. N68335-11-C-0011), and partially
by the U.S. Army Research Office through a Multidisciplinary University Research Initiative (Grant No. W911NF-09-1-0539).
The authors are grateful to Daniel Smith (COMSOL Inc.) for useful discussions relating to numerical ray-tracing algorithms, and to Nathan Kundtz (Intellectual Ventures) for his comments on the utility of conformal maps in cloaking devices.

\end{document}